\documentclass[preprint,5p,twocolumn]{elsarticle}
\usepackage{hyperref}
\usepackage{lineno}
\usepackage{upgreek}
\usepackage{float}
\usepackage{bm}
\usepackage{xcolor}
\usepackage{ulem}
\usepackage{amsmath}
\usepackage{graphicx}
\usepackage[english]{babel}
\journal{Chemical Physics}

\begin{document}
 
\title{Influence of molecular shape on self-diffusion under severe confinement: A molecular dynamics study}

\author[fn1]{I. Dhiman}
\author[fn4]{U. R. Shrestha}
\author[fn2]{D. Bhowmik\corref{mycorrespondingauthor1}}
\cortext[mycorrespondingauthor1]{Corresponding author}
\ead{bhowmikd@ornl.gov}
\author[fn3]{D. R. Cole}
\author[fn3]{S. Gautam}

\address[fn1]{Neutron Scattering Division, Oak Ridge National Laboratory, Oak Ridge, Tennessee 37831, USA.}
\address[fn4]{Center for Molecular Biophysics, University of Tennessee/Oak Ridge National Laboratory, Oak Ridge, Tennessee 37831, USA.}
\address[fn2]{Computational Science and Engineering Division, Oak Ridge National Laboratory, Oak Ridge,  Tennessee 37831, USA.}
\address[fn3]{School of Earth Sciences, The Ohio State University, 275 Mendenhall Laboratory, 125 S Oval Drive, Columbus, Ohio, USA.}

\date{\today}

\begin{abstract}

We have investigated the effect of molecular shape and charge asymmetry on the translation dynamics of confined hydrocarbon molecules having different shapes but similar kinetic diameters, inside ZSM-5 pores using molecular dynamics simulations. The mean square displacement of propane, acetonitrile, acetaldehyde, and acetone in ZSM-5 exhibit two different regimes - ballistic and diffusive/sub-diffusive. All the molecules except propane exhibit sub-diffusive motion at time scales greater than 1 ps. The intermediate scattering functions reveal that there is a considerable rotational-translational coupling in the motion of all the molecules, due to the strong geometrical restriction imposed by ZSM-5. Overall the difference in shape and asymmetry in charge imposes severe restriction inside the ZSM-5 channels for all the molecules to different extents. Further, the behavior of molecules confined in ZSM-5 in the present study, quantified wherever possible, is compared to their behavior in bulk or in other porous media reported in literature.

\end{abstract}

\maketitle

\section*{Notice of Copyright}
{Manuscript has been authored by UT-Battelle, LLC under Contract No. DE-AC05-00OR22725 with the U.S. Department of Energy. The United States Government retains and the publisher, by accepting the article for publication, acknowledges that the United States Government retains a non-exclusive, paid-up, irrevocable, worldwide license to publish or reproduce the published form of this manuscript, or allow others to do so, for United States Government purposes. The Department of Energy will provide public access to these results of federally sponsored research in accordance with the DOE Public Access Plan (http://energy.gov/downloads/doe-public-access-plan)}

\section{\label{Intro} Introduction}

Understanding molecular motion under geometrical restrictions imposed by porous  media, such as zeolites, is important from both an application as well as fundamental physics viewpoint \cite{Cole, Ruthven,Karger1, Buntkowsky}. Zeolites are porous crystalline materials, that can absorb several molecules and are extensively used for industrial applications, such as molecular separation and catalysis, owing to their uniform pore distribution, high void volume space and shape selectivity \cite{Liu, Zheng, Deamer, Sun, Ajayan, Ajayan1}. In particular, petroleum industry cracking of hydrocarbons inside zeolite pores has been of immense importance \cite{Gautam, Tomlinson, Smit}. These applications in turn are dependent upon the translational diffusive properties of the molecules adsorbed in the pores. The diffusive behavior of hydrocarbon in zeolites is determined by different factors such as guest concentration, temperature and dimensions of both hydrocarbon molecules and pores. As a result, it becomes imperative to perform a comprehensive study of the dynamical behavior of hydrocarbons confined in porous materials like zeolites \cite{Wang, Cole1, gautam0, Lash, Snow, Cole2}. In that respect it is crucial to study both the translational and rotational dynamics in order to understand the influence of geometrical confinement on hydrocarbon behavior in zeolite pores.  

It is well known that the translation motion is strongly influenced under confinement. Jobic et al. carried out quasielastic neutron scattering (QENS) measurements to study the translation diffusion of molecules in 1-dimensional channel systems, such as AlPO$_4$-5 and ZSM-48 \cite{Jobic}. With increase in loading single file diffusion is observed for both systems. This single file diffusion behavior has a significant effect on catalytic reaction rates. K$\ddot{a}$rger et al. used Monte Carlo simulations to understand and predict correlations between single file diffusion and reactions \cite{Karger}. Similarly, various QENS and molecular dynamics (MD) simulation based studies have  been reported on widely used Na-Y zeolite, to understand the diffusive behavior of hydrocarbons confined in this zeolite \cite{Sayeed, Gautam1, Mukhopadhyay, Mitra2, Gautam2}. In particular, hydrocarbons such as, propane, acetylene, 1, 3 -butadiene and propylene are approximately three times larger in size as compared to the pore size in Na-Y zeolite \cite{Gautam1, Mukhopadhyay}. The translation dynamics are found to occur at three different time scales. For acetylene, the fastest time scales corresponds to 'free-particle' motion, whereas for other molecules, even the fastest component represents 'diffusive motion'. 
 
In contrast to Na-Y zeolite, ZSM-5 zeolite pores impose stronger restrictions on the confined molecules with pore dimensions of 0.55 nm. The influence of this stronger confinement has been highlighted in several experimental as well as theoretical studies reported in the literature. For example, a number of QENS studies have addressed the dynamical behavior of cyclohexane, benzene, and methanol molecules adsorbed in ZSM-5 \cite{Sahasrabudhe, Tripathi, Mitra, Mitra1}. Further, MD simulation studies on the dynamics of molecules such as xenon, methane, and n-alkanes (\textit{n}-C$_4$ to \textit{n}-C$_{20}$) in silicate have also been reported \cite{Pickett, Demontis, Runnebaum}. The molecular size of cyclohexane and benzene is significantly larger than the pore size of ZSM-5. As a result, it restricts the translation motion and hence only the rotational motion is observed. Interestingly, despite the fact that the molecular size of methanol is smaller than the pore size of ZSM-5, no dynamical activity is observed. This behavior can be attributed to the strong electronic binding of the molecule to the framework, thus constraining the molecular motion. The strong sorption of methanol in ZSM-5, in turn, is a result of the high charge asymmetry (polarity) of the former which highlights the role played by charge asymmetry of the confined molecule in determining its mobility. Additionally, these studies have also shown that the molecular symmetry plays an important role towards understanding the dynamical properties of the adsorbed molecules. A comparative QENS and MD simulation study on propylene confined in Na-Y and Na-ZSM-5 zeolites showed contrasting results \cite{Sharma}. Propylene confined in ZSM-5 zeolite exhibits anisotropic behavior, while in Na-Y no directional dependence is observed in translational diffusion.  This comparative study also shows that translation diffusivity of propylene is slower in ZSM-5 than in Na-Y. This behavior can be correlated with higher geometrical constraint and stronger interaction between ZSM-5 and propylene, (as revealed by an FTIR study \cite {FTIR}) as opposed to that between Na-Y and propylene molecule.

As described earlier, one important aspect that influences the properties of confined fluids is the shape of the confined molecule. Various studies have been reported in literature highlighting the importance of molecular shape. As an example, Schenk et al. have demonstrated that molecules having shapes commensurate with the sieve pores used for catalytic conversion reactions of alkanes promote the formation of reaction intermediaries \cite{Schenk}. The molecular shape effect on the solute transport processes has also been discussed by Van der Bruggen et al. \cite{Bruggen, Bruggen1}, Bhowmik et al. \cite{Bhowmik, Bhowmik1, Bhowmik6} and Santos et al. \cite{Santos}. Similarly, Thalladi et al. have studied the influence of molecular geometry on the melting point of n-alkanes \cite{Thalladi}. 

Here, we report an investigation of translational motion of confined hydrocarbon molecules having different shapes but similar kinetic diameters inside ZSM-5 pores using molecular dynamics simulation. 
Propane, in contrast to all the other molecules studied here, exhibits an absence of charge asymmetry. Acetonitrile displays linear shape and high degree of charge asymmetry. Acetaldehyde and propane have somewhat similar shape, although acetaldehyde shows charge asymmetry. Acetone has a globular shape with charge asymmetry.
These four different molecules have been selected in such a way that the same TraPPE-UA \cite{Martin} formalism can be used to model them. Properties of these molecules are summarized in Figure \ref{fig:ZSM_schematic}. We observed that propane exhibits least restricted translational diffusion among all the studied molecules inside ZSM-5 due to its shape and apolar nature. However, the dynamics slows down with increase in loading due to crowding effect. Conversely, acetonitrile, acetaldehyde and acetone demonstrate highly restricted motion with sub-diffusive behavior due to their  shape and net charge asymmetry.

\section{\label{simudetails} Simulation Details}\

\begin{figure}[H]
\centerline{\includegraphics[height=10.0cm]{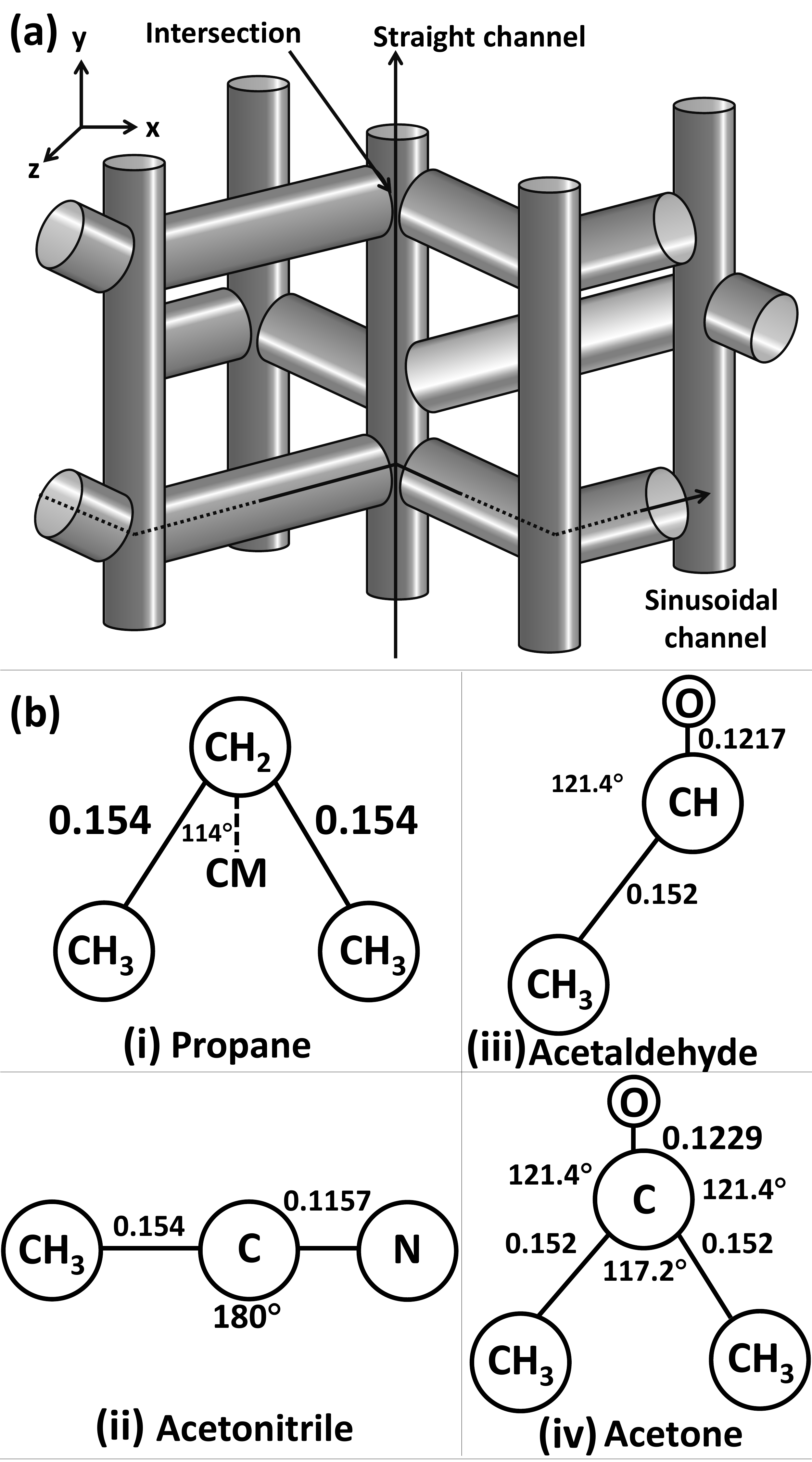}}
\caption{(a) Approximate schematic of ZSM-5 structure. Different channel structures have been indicated, as straight (vertical line), sinusoidal (zig-zag line), and the intersection channels, (b) Schematic of (i) propane, (ii) acetonitrile, (iii) acetaldehyde and (iv) acetone is shown. The bond angles here are in degrees, the bond lengths are in nm units. CM in the propane schematic stands for center of mass. We define the molecular axis along the vector joining the center of mass of the molecule with the reference site (reference sites for Propane:CH$_2$, Acetaldehyde:CH, Acetonitrile:CH$_3$ and Acetone:CH$_3$).}
\label{fig:ZSM_schematic}
\end{figure}

Classical molecular dynamics (MD) simulations have been carried out using DL-POLY 1.9 \cite{Todorov}. The behavior of a series of guest molecules having similar sizes (shown in Figure \ref{fig:ZSM_schematic}(b)) as a function of loading (\textit{n} = 2, 4, 6 and 8 guest molecules per unit cell (mpuc)) in all silica analogue of ZSM-5 (silicalite) have been investigated. Figure \ref{fig:ZSM_schematic}(a) shows an approximate schematic of ZSM-5 in 3D. ZSM-5 structure comprises straight channels along the crystallographic axis \textit{b} (Cartesian direction Y) with $\approx$ 0.55 nm diameter, while sinusoidal channels lie in X-Z plane, which is perpendicular to Cartesian direction Y. At the intersection of these straight and sinusoidal channels a slightly larger pore space of about 0.8 nm exists. 

The guest hydrocarbon molecules are treated using  united-atom and semi-flexible models following TraPPE-UA conventions \cite{Martin}. The TraPPE-UA formalism allows transferability of the force-field parameters and precision in the prediction of thermodynamic properties. Simultaneously, the treatment of H atoms together with other atoms to form pseudo atoms greatly reduces computational demand. TraPPE-UA force field and Lorentz-Berthelot mixing rules have been employed to compute non-bonded coulombic interactions, which determines the Lennard-Jones (LJ) potentials for cross-terms \cite{Allen}. The initial simulation box is constructed by placing the required number of molecules as per the loading in a ZSM-5 unit cell, built a priori using experimentally obtained coordinates \cite{Koningsveld}.

The simulation cell consists of 2$\times$2$\times$3 unit cells of ZSM-5, along all directions, with final cell dimensions of 40.044 $\AA$ $\times$ 39.798 $\AA$ $\times$ 40.149 $\AA$. After setting up the initial model configurations, their energy is minimized for 20 ps at a lower temperature (30 K). Thereafter, each system was allowed to reach its respective equilibrium state by running an NVT simulation for 500 ps with 1 fs time step at 300 K. The system equilibration can be affirmed when fluctuations in temperature and energies (potential and kinetic) fall below 5\%. Periodic boundary conditions are implemented in all directions and long-range interactions are calculated by 3D Ewald sum. Following the TraPPE-UA convention 14 $\AA$ is taken as the cut-off distance for all the interactions. All ZSM-5 atoms remain rigid during the entire simulation. Kopelevich and Chang have shown that the dynamical properties of ethane in silicalite are influenced by lattice vibrations only along some diffusion paths \cite{Kopelevich}. Similarly, for methane in silicalite, Demontis et. al. found that the effect of using a flexible network on self-diffusivity was negligible \cite{flexframe}. Also, maintaining the substrate rigidity has been applied in various similar studies investigating the dynamical properties of confined ethane \cite{Bhide, Gautam4}, propylene \cite{Gautam} and ethene \cite{Jianfen}. We therefore exclude ZSM-5 lattice vibrations in our simulations to facilitate cost-efficient computation. Once equilibration is attained, the simulation production run is continued for another 1.5 ns in NVT ensemble recording the trajectories at an interval of 0.02 ps post-simulation analysis.

\section{\label{results} Results}

\subsection{\label{distribution} Molecular Distribution}\

\begin{figure}[H]
\centerline{\includegraphics[height=8.2cm]{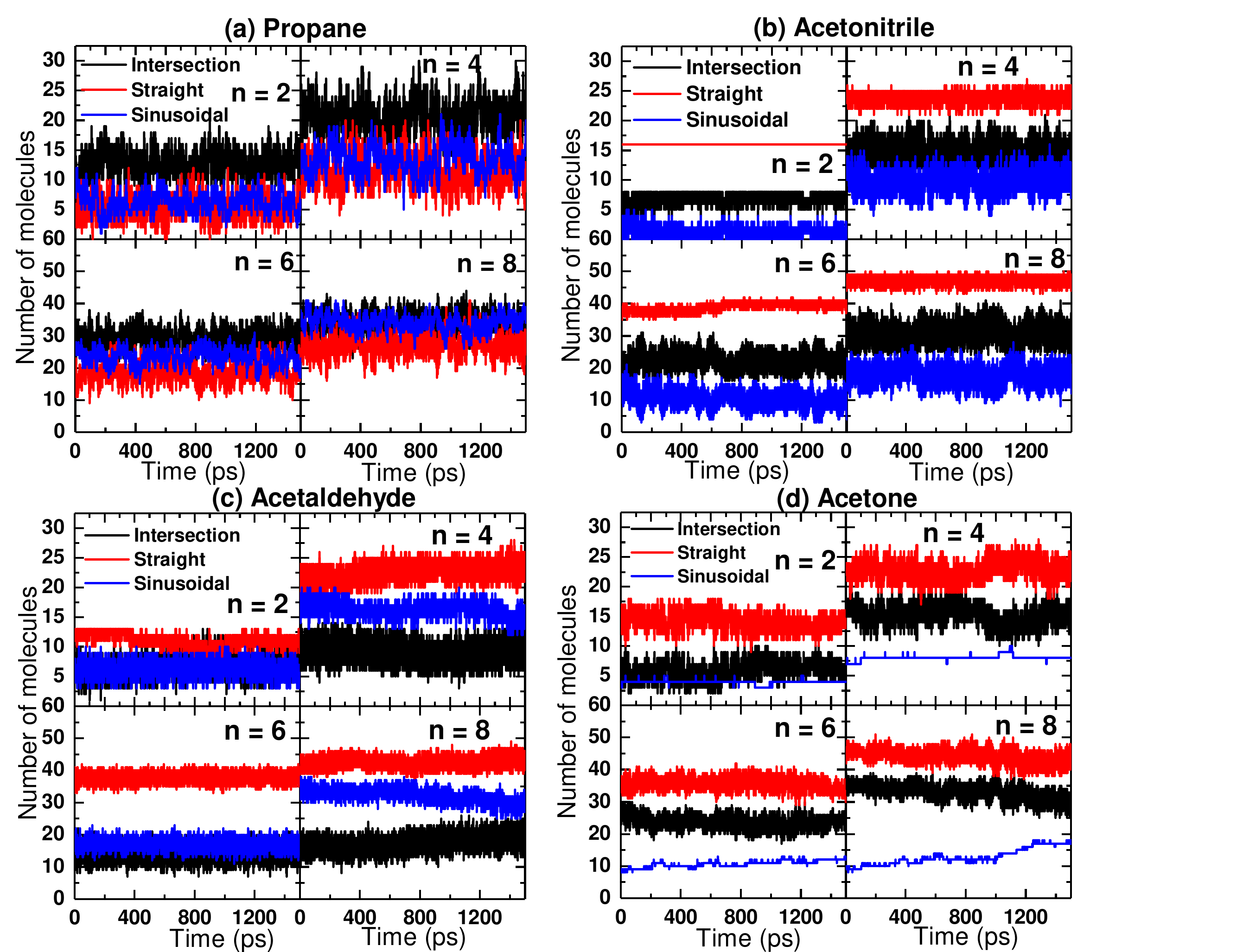}}
\caption{Distribution of (a) propane, (b) acetonitrile, (c) acetaldehyde and (d) acetone molecule in different channels, straight, sinusoidal and intersection in ZSM-5.}
\label{fig:Distribution_all}
\end{figure}

As ZSM-5 consists of pore spaces of different shapes (straight and sinusoidal channels and ellipsoidal intersections), it provides an opportunity to study the effect of pore variability in the same system \cite{Gautam4}. For example, we can test to see if the guest molecules show a preference for a given pore type over the others. To study the distribution of molecules in different channels of ZSM-5, the number of molecules as a function of time is plotted in Figure \ref{fig:Distribution_all} for all the molecules at various loadings. The absolute value defines the number of molecules at a particular instance in a given pore or channel, while the fluctuations indicate the exchange of molecules with other possible channels. No fluctuation (i.e. a solid straight line in Figure \ref{fig:Distribution_all}) would suggest that there is no exchange of molecules from that channel. Among all the molecules only propane has no charge asymmetry. Figure \ref{fig:Distribution_all}(a) shows a significant mutual exchange of propane molecules between different channels of ZSM-5. It can be clearly observed that the number of molecules is greatest in the intersection region. Owing to a slightly larger diameter at intersections (0.8 nm) we see a higher number of propane molecules at lower loading ({\textit{n}} = 2 mpuc) in the intersection. However, with an increase in loading there is gradually less space available for the molecules in intersections and they start populating other channels.

For acetonitrile (Figure \ref{fig:Distribution_all}(b)) (\textit{n} = 2 mpuc), there is no exchange of molecules from the straight channel. The number of acetonitrile molecules is the highest in this channel - possibly indicating that the linear structure is better accommodated in straight channel at \textit{n} = 2 mpuc loading. On the other hand, there is a significant mutual exchange of acetonitrile molecules between the intersection and the sinusoidal channels as evident from the fluctuations at this loading. Small number of acetonitrile molecules in the sinusoidal channels at all loadings indicate that this channel is preferred least perhaps because of the tortous structure of this channel. With increase in loading there is no difference noted in the general behavior, except for some small exchange of molecules with straight channels. With increasing loading more and more molecules begin to enter the other two regions as well. A similar behavior is expected for acetaldehyde and acetone, but this is not the case (Figures \ref{fig:Distribution_all}(c) and \ref{fig:Distribution_all}(d)). Unlike acetonitrile, these two molecules show some degree of exchange of molecules with others, irrespective of the loading. Acetone in some ways follows the trend of acetonitrile where straight and sinusoidal channels are the most and least favored channels, respectively. However, we also see the sinusoidal channels are not favored in exchanging molecules, which is not the case for acetronitrile. For acetaldehyde straight channels are the most populated, while the intersections are the least populated. One can also note that number of molecules can gradually become altered (increase in acetone and decrease in acetaldehyde) in sinusoidal channels with time. The behavior of propane is different from the other molecules because it tends to exhibit a slight preference for intersection, whereas all other molecules show a stronger preference for straight or sinusoidal channels.

\subsection{\label{dynamics} Translational motion}\

Mean square displacement (MSD) of the center of mass is calculated to understand the translational diffusive motion of the molecules in ZSM-5. Time dependence of the MSD reveals information about the nature of translational dynamics. For pure diffusive behavior MSD should exhibit linear dependence as a function of time while in case of sub- or super- diffusive motion MSD departs from linear behavior. 

\begin{figure}[H]
\centerline{\includegraphics [height=7.8 cm]{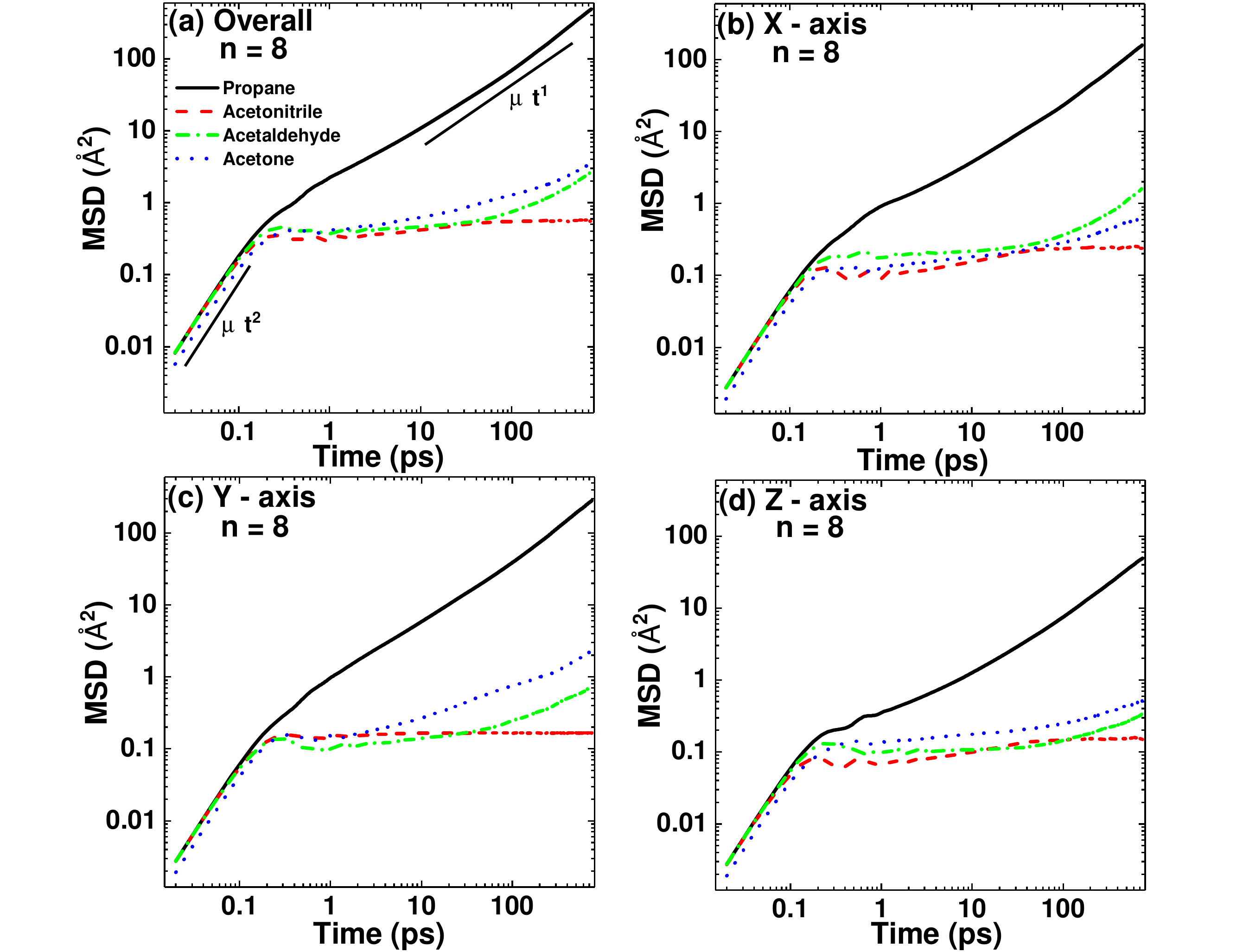}}
\caption{Mean squared displacement (MSD) for propane, acetonitrile, acetaldehyde and acetone in ZSM-5 (a) Overall, and with respect to X- (b), Y- (c) and Z- (d) direction for \textit{n} = 8 mpuc. The two different time regimes are highlighted in the Figure \ref{fig:MSD_combined}(a) for propane.}
\label{fig:MSD_combined}
\end{figure}

In Figure \ref{fig:MSD_combined}, the temporal variation of total MSD of all the molecules is shown along with their components in X-, Y- and Z-directions for \textit{n} = 8 mpuc. The time dependence for all the other loadings, \textit{n} = 2, 4 and 6 mpuc, exhibits similar behavior (see supplementary material for MSD plots at loadings other than 8 mpuc). The Y component of MSD exhibits the highest value, in comparison to X- and Z- components. This is a direct consequence of ZSM-5 pore structure, wherein the straight channels lying along Y-axis provide opportunity for one dimensional motion of the molecules. On the other hand, relatively higher geometrical restriction is imposed by the tortuosity of the sinusoidal channels along X- and Z-directions, respectively.

Closer inspection of the MSD results reveals two distinct time regimes with different behaviors. As an example, these two regimes with different behaviors are highlighted in the Figure \ref{fig:MSD_combined}(a) for propane at \textit{n} = 8 mpuc. Initially, in the first regime (MSD $\propto$ (\textit{t})$^2$) upto $\approx$ 0.3 ps MSD shows a sharp increase with time. This regime corresponds to ballistic motion of the molecules that move freely without interacting with the neighboring molecules. Also owing to an absence of interactions with neighboring molecules, no significant change in this ballistic regime is observed as a function of loading. For time scales greater than 1 ps, the MSD shows linear time dependence (MSD $\propto$ (\textit{t})$^1$), implying diffusive motion. MSD displays a continuous decrease as a function of loading,  indicating reduced diffusivity at higher loading values. 

Unlike propane, acetonitrile exhibits a different time dependent behavior of MSD, shown in Figure \ref{fig:MSD_combined} for \textit{n} = 8 mpuc. After the first ballistic regime, at longer time scales above 1 ps, no significant variation in MSD is observed in all three directions, wherein the trends display a plateau. Also, in comparison to propane, the MSD values are significantly reduced by nearly 3 orders of magnitude, implying that acetonitrile does not move a very large distance and is severely restricted. This indicates the motion is strongly sub-diffusive. This sub-diffusive behavior also known as anomalous diffusion is a signature characteristic of polymers \cite{Bhowmik2, Bhowmik3, Bhowmik4} and bio-macromolecules like protein \cite{Shrestha} or RNA \cite{Bhowmik5} in crowded environments. In the present case, this behavior results from a crowded environment, that inhibits molecules from passing each other.
For acetaldehyde, the time evolution behavior exhibits three different regimes. The initial ballistic regime is followed by a strongly sub-diffusive behavior, and at much higher time scales, it tends towards being diffusive. Also, MSD values similar to acetonitrile are obtained, indicating a restrictive motion. The time dependence of MSD for acetone, as depicted in Figure \ref{fig:MSD_combined}, also shows signatures of anisotropic and restrictive motion of molecules. Although, higher MSD for acetone in comparison to acetonitrile and acetaldehyde shows less restrictive motion of acetone molecules in ZSM-5. In comparison to propane, acetone still exhibits almost an order of magnitude lower MSD. Again for acetone, three different time regimes in MSD versus time plot are observed, with ballistic, sub-diffusive and slightly enhanced motion at longer time-scales. To investigate if this long time motion might eventually turn diffusive for time greater than 1.5 ns we carried out longer simulations lasting 14 ns for acetaldehyde, acetonitrile and acetone at \textit{n}=8 mpuc. No diffusive motion was observed even at longer times for any of these systems (Figure S1).   

\begin{figure}
\centerline{\includegraphics[height=7.5cm]{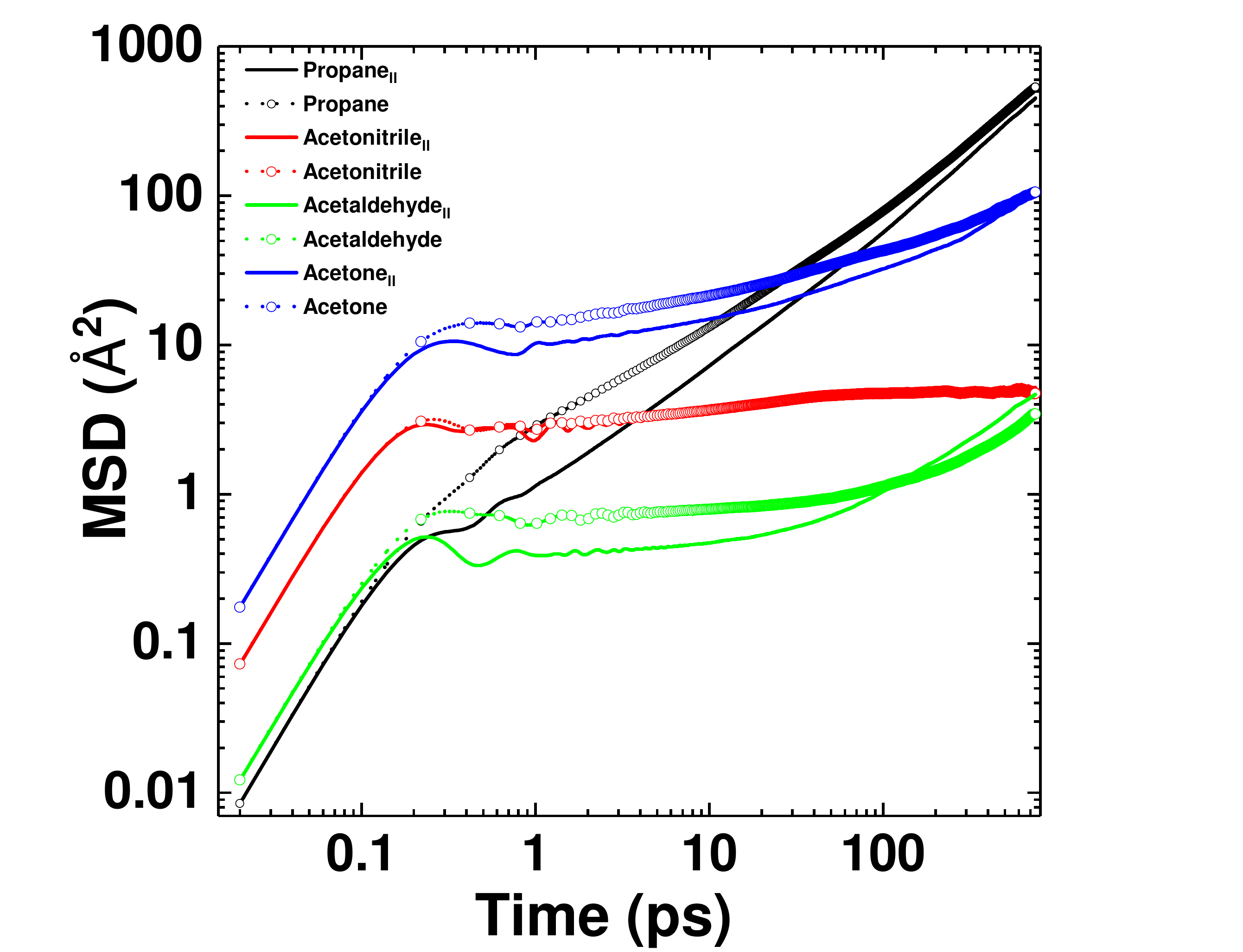}}
\caption{Mean squared displacement (MSD) for all the molecule in ZSM-5 in the directions parallel and perpendicular to the molecular axis for \textit{n} = 8 mpuc.}
\label{fig:MSD_Parallel_Perpendicular_combined}
\end{figure}

For a linear molecule like acetonitrile, motion along the direction of the molecular axis can be expected to be easier, especially when the molecule is confined to motion in a channel. To explore this further, we calculated the components of MSD in a direction parallel to molecular axis (see Figure 1 (b)) for all molecules. In addition, the MSD in a plane perpendicular to the molecular axis is also calculated. The MSD resolved in these directions and planes for all molecules are shown in Figure \ref{fig:MSD_Parallel_Perpendicular_combined}. Acetonitrile does not exhibit any preference in the direction of motion in the molecular frame although other molecules do seem to exhibit a slight preference for motion perpendicular to the molecular axis at times below 100 ps. Although this is counterintuitive considering that acetonitrile is a linear molecule, considering the overall MSD in Figure \ref{fig:MSD_combined}, it can be seen that the motion of this molecule is severely restricted and is limited to a root mean squared displacement of less than even its molecular size. This means that the motion of acetonitrile is more like shuttling in a small, roughly spherical region, rather than a one dimensional long range motion as is expected in ZSM-5 channels. For other molecules, the slightly preferred motion perpendicular to the molecular axis results from the arbitrary definition of molecular axis for a globular, rather than a linear molecule.

\begin{figure}
\centerline{\includegraphics[height=8.0cm]{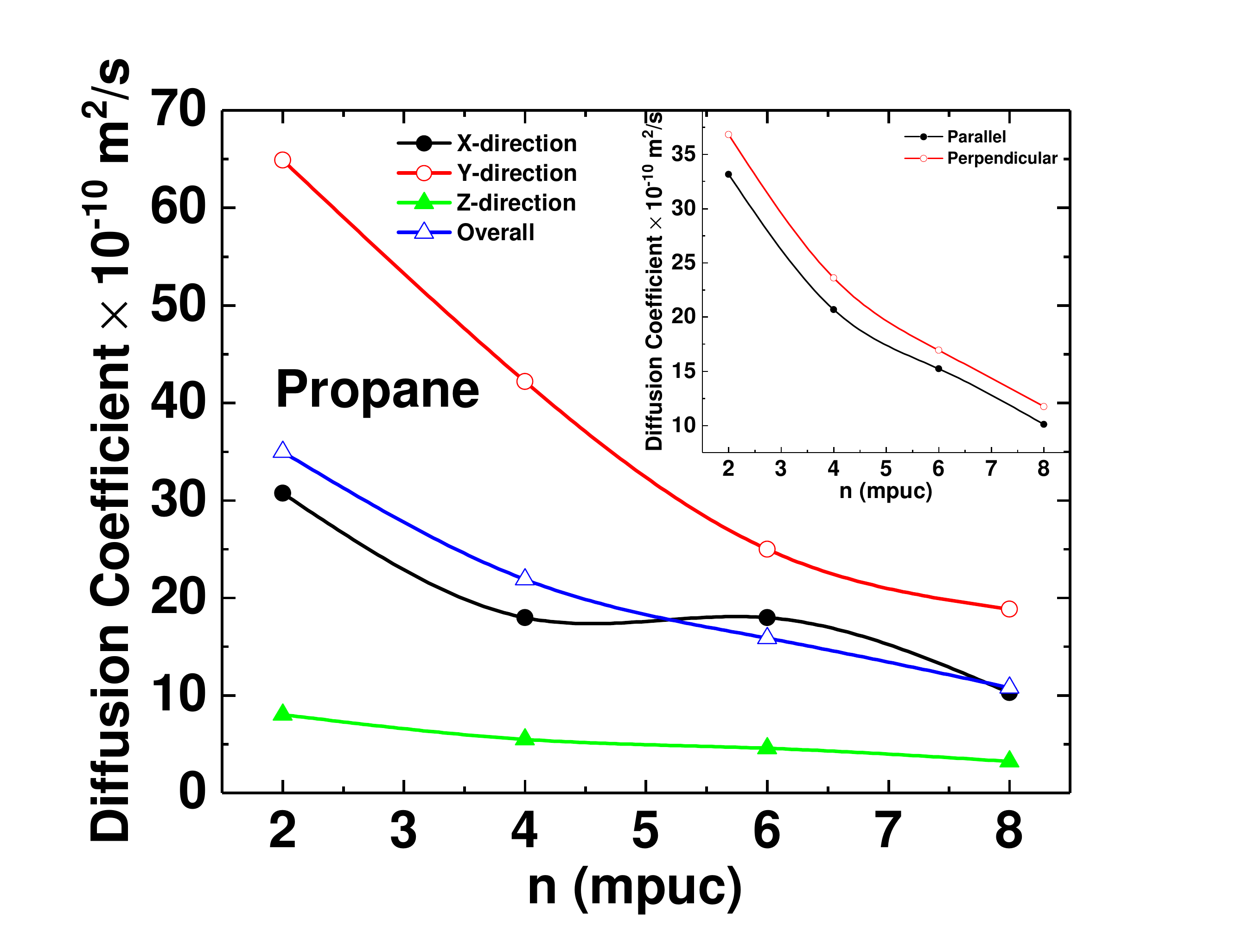}}
\caption{Diffusion coefficient of propane along X-, Y- and Z-direction as a function of loading. The inset to the figure shows the diffusion coefficient of propane in the direction parallel and perpendicular to molecular axis.}
\label{fig:diffusioncoeff_propane}
\end{figure}

Diffusion coefficients (D) in X-, Y-, and Z- direction and in the directions parallel and perpendicular to the molecular axis for propane are calculated as a function of loading ($D=\frac{MSD}{2n_{d}t}$ with $n_{d}$ being the dimensions) when time scale is longer than $\approx$ 1 ps, satisfying the linear behavior of MSD required for diffusive motion. As no diffusive regime is observed for molecules other than propane, these calculations were limited to the latter and the results are plotted in Figure \ref{fig:diffusioncoeff_propane}. As expected, the diffusion coefficient exhibits a maximum in the Y- direction, while minimum in the Z- direction. Also, with increasing loading the diffusion coefficient is decreased, due to the crowding of molecules at higher loading. Additionally, along directions parallel and perpendicular to the molecular axis, diffusion coefficients (D) exhibit a decrease as a function of loading. 

Self intermediate scattering function (ISF), \textit{I(Q,t)}, is useful to obtain the spatial-temporal information about the molecular motion. The \textit{I(Q,t)} is the Fourier transform of the van-Hove correlation function \textit{G(\textbf{r}, t)}. Classically, this function gives the probability of finding a molecule at position '\textit{\textbf{r}}' at time '\textit{t}', provided that same molecule was present at origin at \textit{t} = 0. ISF can be measured in inelastic neutron scattering experiments. In MD simulations, the ISF is obtained from the particle trajectories as, 

\begin{equation}
\textit{I(Q,t)} =\left \langle exp\left \{ i\textbf{\textit{Q}}.[\textbf{\textit{r}}_{i}(t+t_{0})-\textbf{\textit{r}}_{i}(t_{0})] \right \} \right  \rangle
\label{ISF}
\end{equation}

where \textbf{\textit{Q}} is the wave-vector transfer, related to the angle of scattering in the experiment, $\textbf{\textit{r}}_{i}(t+t_{0})$ and $\textbf{\textit{r}}_{i}(t_{0})$ are the position vectors of the \textit{i$^{th}$} particle at times \textit{t + t$_{0}$} and \textit{t$_{0}$}, and the angular brackets indicate an ensemble average. An average over different \textbf{\textit{Q}} vectors with the same magnitude is also taken on the right hand side of the equation. Further, to extract information about the translational motion, the \textit{r$_{i}$} in the above equation can be  written as \textbf{\textit{r}}$_{i}$ = \textbf{\textit{r}}$_{i}^{C. M.}$ + \textbf{\textit{d}}$_{i}$, where \textbf{\textit{r}}$_{i}^{C. M.}$ represents the position vector of the center of mass in the space fixed frame of reference and \textbf{\textit{d$_{i}$}} is the particle position vector with respect to the center of mass. Translational component of motion can be studied by following the evolution of \textbf{\textit{r}}$_{i}^{C. M.}$, whereas the evolution of \textbf{\textit{d$_{i}$}} can be used to study roto-vibrational motion.
Therefore, the translational intermediate scattering function (TISF) (\textit{I$^{C.M.}$(Q,t)}) can be written as  

\begin{equation}
\textit{I$^{CM}$(Q,t)} =\left \langle exp\left \{ i\textit{\textbf{Q.}} [\textbf{\textit{r}}_{i}^{C. M.}(t+t_{0})-\textbf{\textit{r}}_{i}^{C. M.}(t_{0})] \right \} \right  \rangle
\label{TISF}
\end{equation}

\begin{figure}
\centerline{\includegraphics[height=7.5cm]{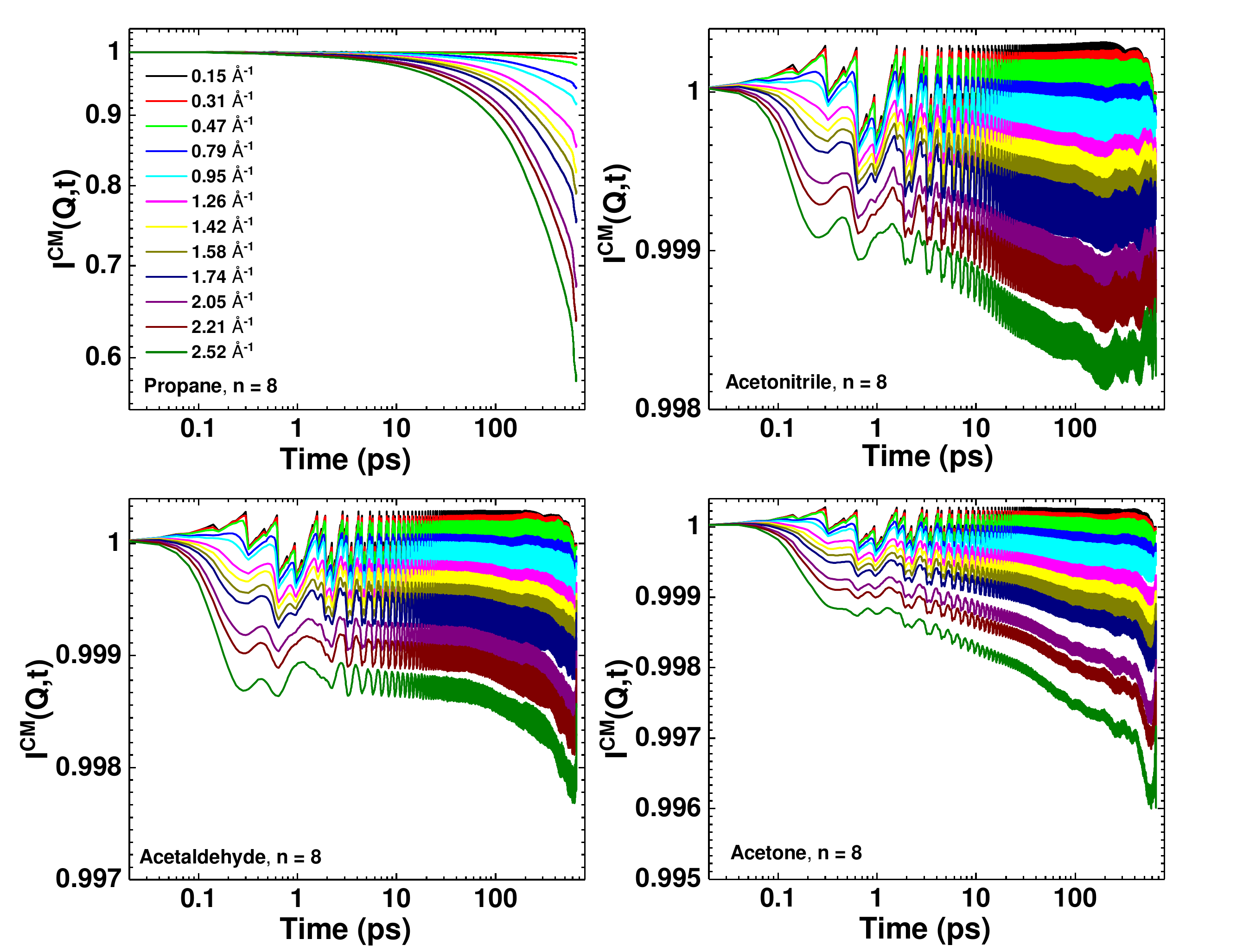}}
\caption{Translational intermediate scattering function, \textit{I{$^{C.M.}$}(Q, t)}, plot for (a) propane, (b) acetonitrile, (c) acetaldehyde and (d) acetone molecule in ZSM-5 for different Q values at \textit{n} = 8 mpuc.}
\label{fig:Combined_Iqt_trans}
\end{figure}

\begin{figure}
\centerline{\includegraphics[height=7.6cm]{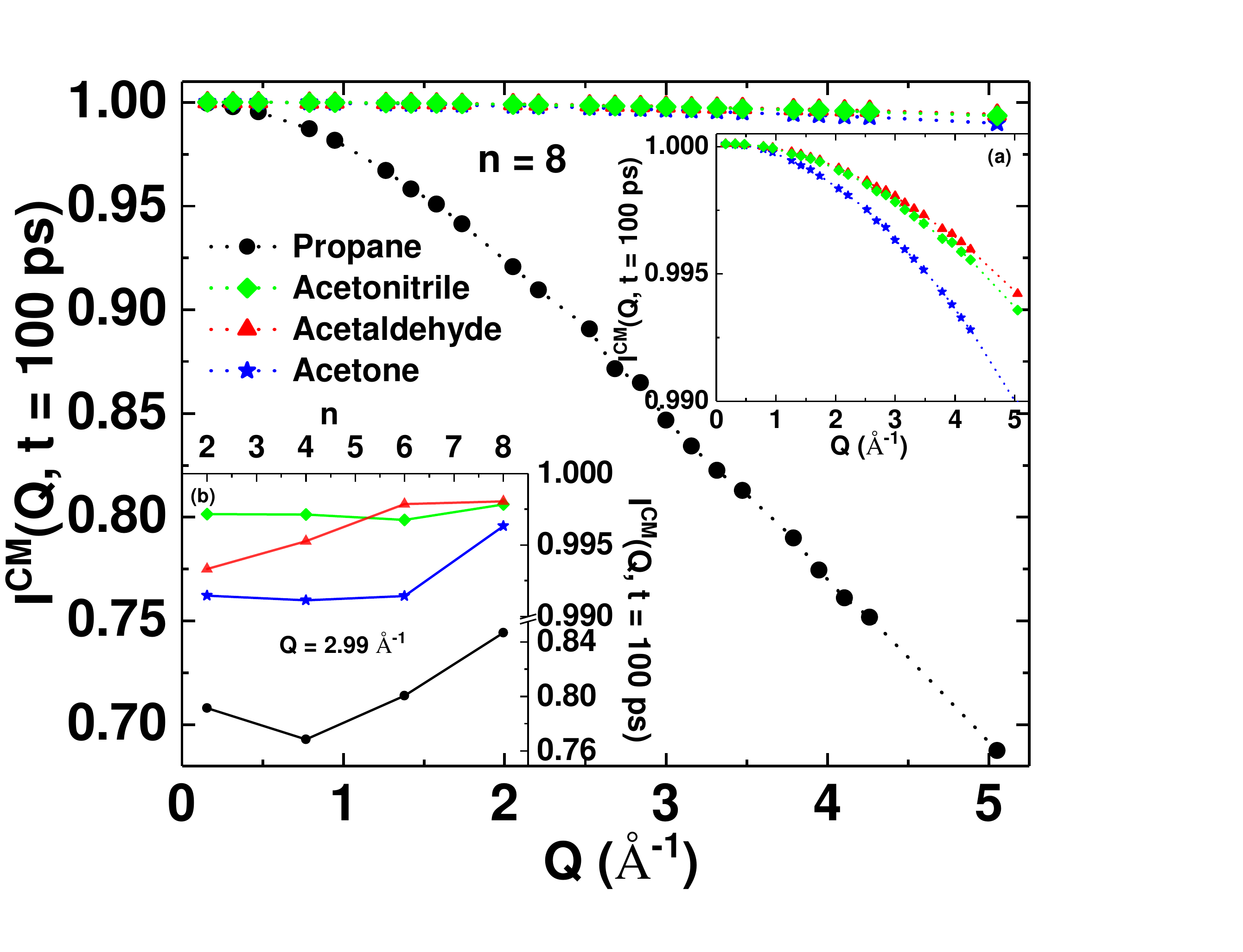}}
\caption{{\textit{I$^{C.M.}$(Q, t = 100 ps)}} as a function of Q for all the molecules. For better clarity, the variation of {\textit{I$^{C.M.}$(Q, t = 100 ps)}}  for acetonitrile, acetaldehyde and acetone molecules is again shown in the inset (a). Inset (b) shows {\textit{I$^{C.M.}$(Q, t = 100 ps)}} at  Q = 2.99 $\AA^{-1}$ as a function of loading. In Figure S2 {\textit{I$^{C.M.}$(Q, t = 100 ps)}} at Q = 0.94 $\AA^{-1}$ as a function of loading is shown.}
\label{fig:Combined_EISF_trans}
\end{figure}

Figure \ref{fig:Combined_Iqt_trans} shows the time dependence of the TISF for all the molecules for \textit{n} = 8 mpuc loading at various \textit{Q} values. This figure mainly represents the translational component of motion, which is probed by replacing the entire molecule by its center of mass. This way, any contribution from rotational or vibrational motion is removed. In a real experiment, the measured functions will have contributions from all different types of motion. In case the rotational and vibrational motions are fast enough to lie outside the experimental window of the instrument used, the functions shown in Figure \ref{fig:Combined_Iqt_trans} will be very close to the measured quantities. For propane in Figure \ref{fig:Combined_Iqt_trans}(a), it can be seen that the TISF start to decay slowly only after a few ps and do not completely decay to zero even after 600 ps. This indicates restriction to molecular motion. In comparison, all the other molecules show very small or no decay at all, implying that relatively greater restriction is imposed by ZSM-5 on acetonitrile, acetaldehyde and acetone. 

In an unrestricted system, given enough time, a typical molecule is expected to cover a large distance such that the correlation between its initial and final positions is lost completely. The TISF for such a system will decay to zero at long enough time. In the present case, however, as the molecular motion is restricted by the confining framework, the TISF attain a non-zero value even at long times. These long time values of TISF, in turn, can be useful in estimating the extent of restriction imposed on the molecular motion. In Figure \ref{fig:Combined_EISF_trans} TISF at t = 100 ps is shown as a function of {\textit{Q}}. Propane can be seen to be least restricted. Further, as \textit{Q} is a reciprocal space quantity that has dimensions of inverse length, the \textit{Q} variation of the TISF facilitates studying motion at different length scales, which is not possible by studying only the behavior of MSD \cite{PCCP}. It can be seen that the TISF attain low values at higher \textit{Q} for propane. However, for other molecules, the \textit{Q} variation is much weaker. This implies that at smaller length scales the motion of propane is progressively less restricted, whereas motion by the other molecules is mostly independent of length scales being probed. This is because for molecules other than propane, the motion is restricted to very small regions as indicated by the very small MSD values. This means no motion can be observed at length scales larger than this region, or in other words at lower \textit{Q} values. Additionally, no significant change in {\textit{I$^{C.M.}$(Q, t = 100 ps)}} as a function of loading is observed for any of the molecules studied in the present work.

\begin{figure}
\centerline{\includegraphics[height=7.5cm]{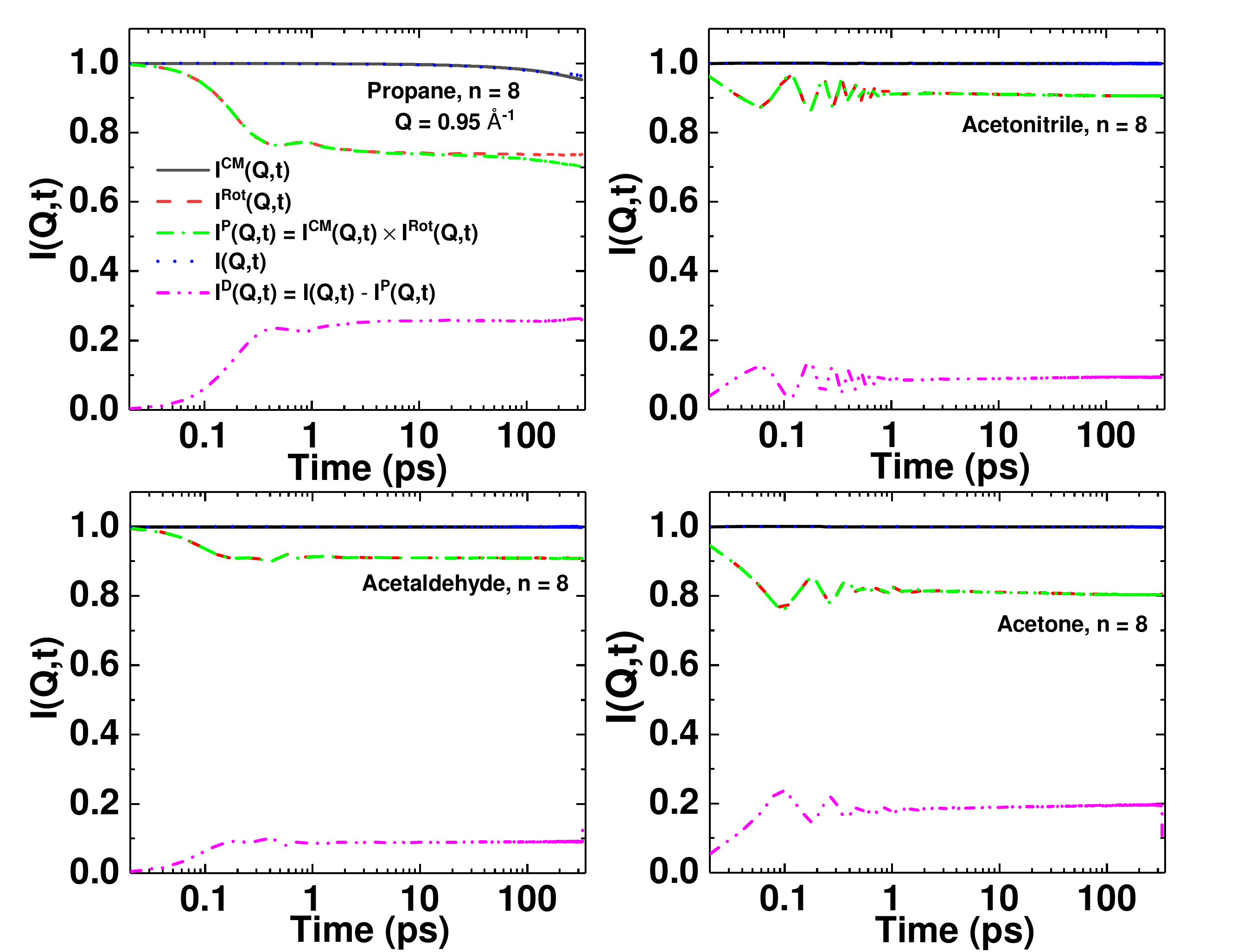}}
\caption{Comparison of different intermediate scattering functions, the center-of-mass/translational motion, rotational motion, product of translational and rotational motion,  net molecular motion (total ISF) and the difference between the net molecular motion and product of translational and rotational motion.} 
\label{fig:Combined_Iqt_comparison}
\end{figure}

An important assumption that facilitates data interpretation in spectroscopy is that different types of motion - translation, rotation, vibration - are independent of each other. This assumption might not always be valid. For some systems these motions might be coupled. For example, Banerjee et al. \cite{coupling} have shown that the rotational and translational motion of ions in aqueous solutions are strongly coupled. To investigate if the assumption of decoupling of motions is valid in our system we have compared the TISF with the ISF for rotational motion as reported in a previous work \cite{Dhiman} and the ISF calculated with co-ordinates of particles in the space fixed frame (i.e. using Eq.{\ref{P_ISF}}). In our system, all three motions, translation, rotation and vibration are present. For the case they are decoupled, the ISF in Eq.{\ref{P_ISF}} can be written as a simple arithmetic product of the ISFs corresponding to the three motions. As vibrational motion is fast and occurs at very short time scales, at longer times its contribution to the net motion can be neglected and we can write ISF in Eq.{\ref{P_ISF}} as the product of TISF and rotational ISF. The difference between the total ISF and the product of the two (considering they are decoupled) this difference would indicate the coupling strength. Therefore, we calculated the product of ISF for rotation motion (taken from ref. \cite{Dhiman}) and translation or center of mass motion. 

\begin{equation}
\textit{I$^{P}$(Q,t)} = \textit{I$^{CM}$(Q,t)}. \textit{I$^{rot}$(Q,t)}
\label{P_ISF}
\end{equation}

In Figure \ref{fig:Combined_Iqt_comparison}, the comparison between the ISF for all the molecules (\textit{n} = 8 mpuc) at \textit{Q} = 0.95 $\AA ^{-1}$ are plotted. In this figure, the ISF for center-of-mass translational motion \textit{I$^{CM}$(Q,t)}, rotational motion \textit{I$^{rot}$(Q,t)}, product of translational and rotational motion \textit{I$^{P}$(Q,t)}, net molecular motion \textit{I(Q,t)} and the difference between \textit{I(Q,t)} and \textit{I$^{P}$(Q,t)} are shown. Substantial differences between \textit{I$^{P}$(Q,t)} and \textit{I(Q,t)} at long times indicate strong coupling between rotational and translational motion. One can also observe that the lesser the restriction on motion (e.g. in propane), the greater the coupling suggesting translational and rotational dynamics are strongly linked.  

\section{\label{Discussion} Discussion}\ 

\begin{figure}
\centerline{\includegraphics[height=15.0cm]{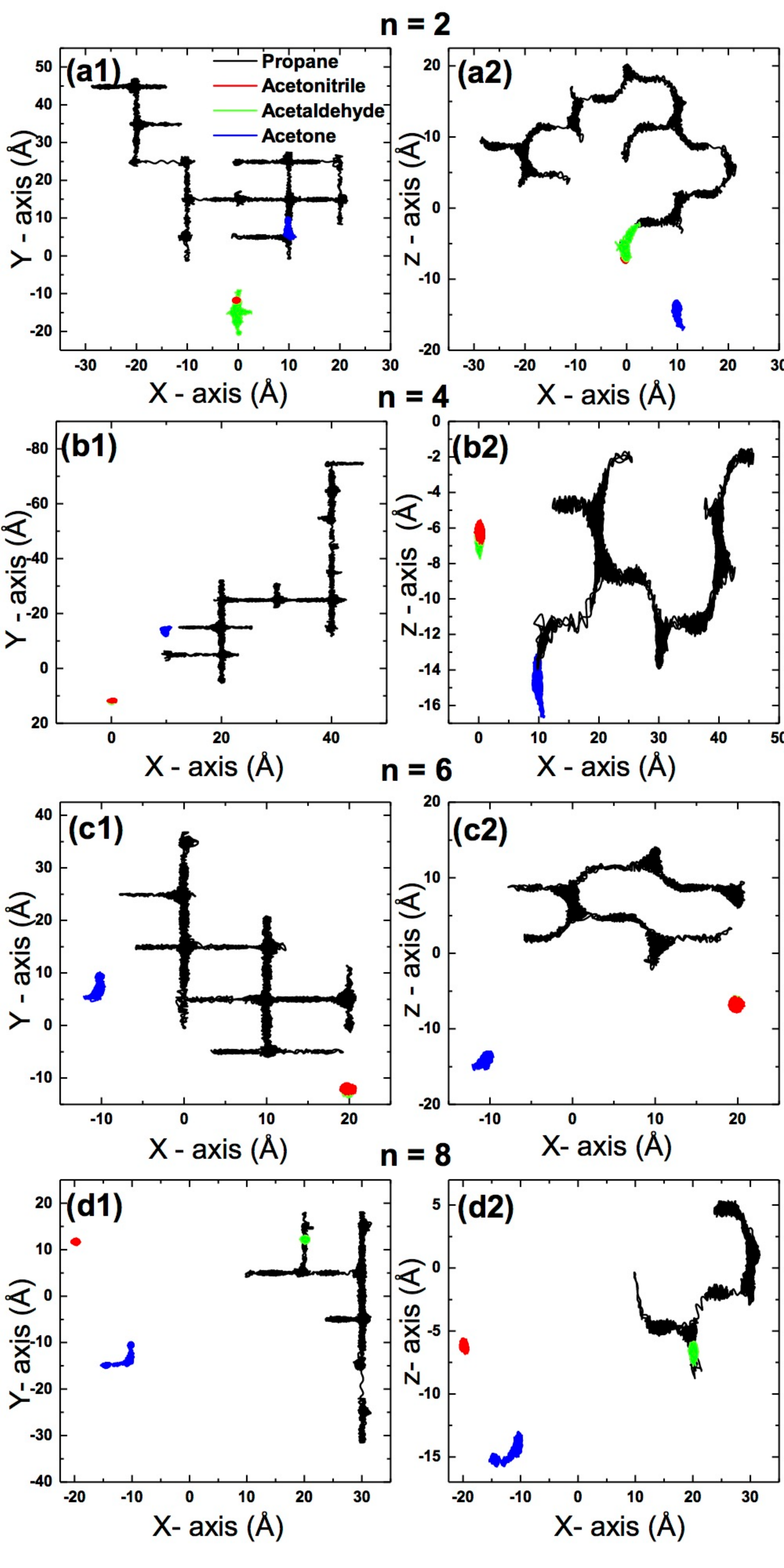}}
\caption{Trajectories of one tagged propane, acetonitrile, acetaldehyde and acetone molecules projected on the X-Y and X-Z plane in ZSM-5 as a function of loadings (a1,a2) \textit{n} = 2, (b1,b2) \textit{n} = 4, (c1,c2) \textit{n} = 6, and (d1,d2) \textit{n} = 8 mpuc.}
\label{fig:trajectories_together}
\end{figure}

\begin{figure}[H]
\centerline{\includegraphics[height=16.5cm]{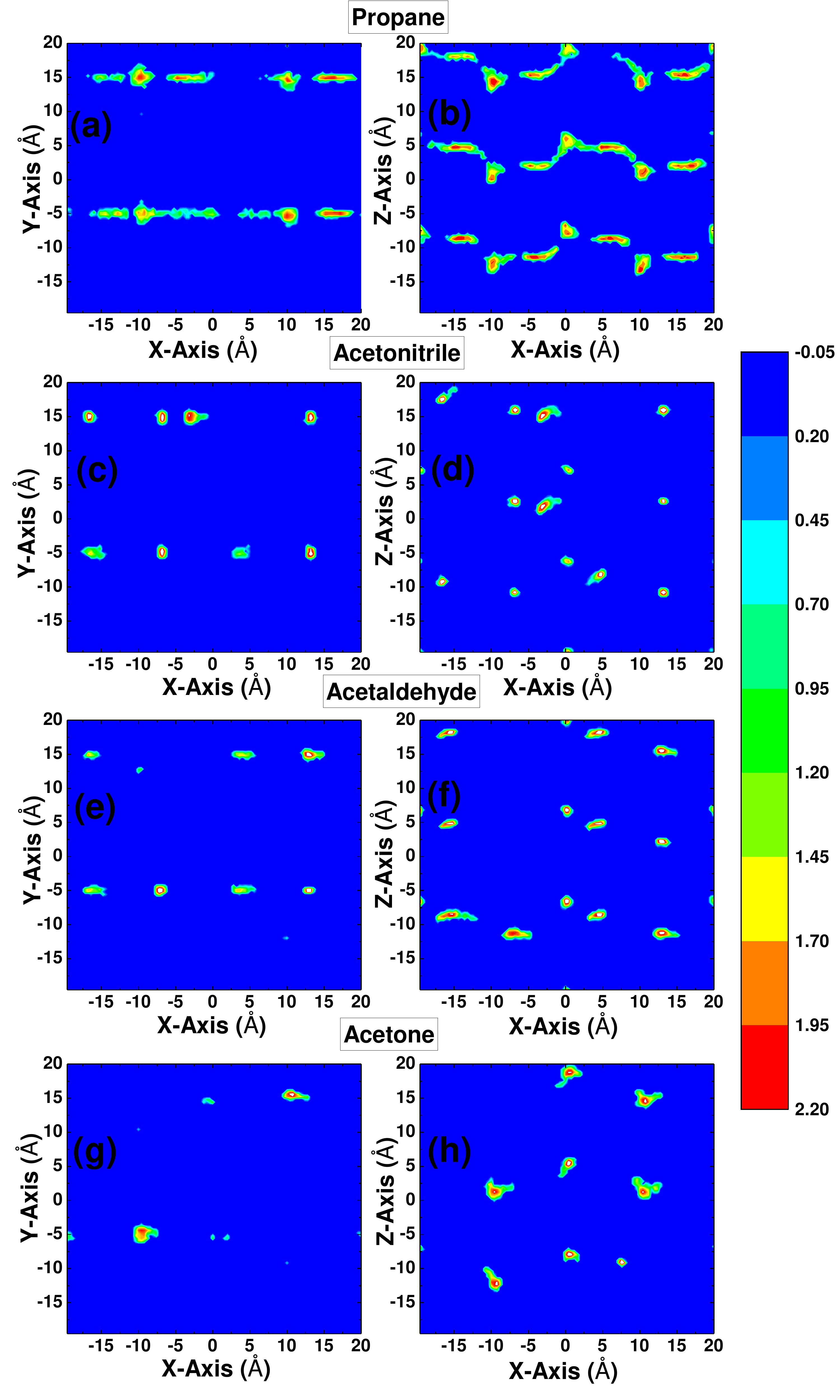}}
\caption{Probability density distribution (in logarithmic scale) of the center of mass of (a, b) propane, (c, d) acetonitrile, (e. f) acetaldehyde and (g, h) acetone in the X-Y and X-Z plane in ZSM-5 for \textit{n} = 8 mpuc.}
\label{fig:Density_All}
\end{figure}

In order to correlate and compare the motion of all the molecules, trajectories of a tagged molecule and the probability density distributions of all the molecules have been plotted in Figures \ref{fig:trajectories_together} and \ref{fig:Density_All}, respectively. The probability distribution function shows the relative probability of finding a molecule at a given location and is calculated for two limited regions of the simulation cell between Y=1.4 nm to 1.8 nm encompassing the entire X-Z plane, and between Z=1.4 nm to 1.8 nm and encompassing the entire X-Y plane.  Distinct differences between different molecules can be observed here. In particular propane  traverses through straight and sinusoidal channels, as expected for several inter-channel migrations (Figure \ref{fig:trajectories_together}). It can also be observed that the trajectory traced by propane intermittently follows the ZSM-5 structure. This is in contrast to all the other molecules, wherein extremely restricted behavior with no or very small movement is observed. This clearly shows the strong influence of confinement on translation/diffusion of different shaped molecules in concert with presence or absence of charge asymmetry. An asymmetric charge distribution makes the guest-host interaction stronger, thereby slowing down the guest molecule. This strengthening of the guest-host interaction gives rise to interesting phenomena. For example, in a study of propane diffusion in silica aerogel, it has been observed that addition of CO$_{2}$ makes the confined propane diffuse faster. The reason for this lies in the stronger interaction of the silica aerogel pore walls with CO$_{2}$ as compared to propane. Because of the non-zero quadrupole moment of CO$_{2}$, it replaces the propane from the pore surface, thus enhancing the propane mobility \cite{Gautam_a}. A similar effect has also been observed in the case of ethane + CO$_{2}$ mixture in controlled pore glass \cite{Gautam_b}. 

The Probability density distributions plotted in Figure \ref{fig:Density_All} exhibit features in agreement with the trajectories plot (Figure \ref{fig:trajectories_together}), particularly for propane. This behavior is also in agreement with the distribution plot shown in Figure \ref{fig:Distribution_all}. The distribution plot shown in Figure \ref{fig:Distribution_all} for all the molecules shows signatures of (strongly) restrictive motion, as a result of confinement imposed by ZSM-5 structure. The distribution of these molecules in different channels of ZSM-5, with straight channels along Y- direction and sinusoidal channels in X- and Z- directions, shows strong dependence on geometrical shape of the molecules. 

In comparison to all the molecules studied here, propane exhibits least restrictive movement in all the ZSM-5 channels. This behavior of propane can be correlated with the shape and absence of charge asymmetry, which imposes least restriction on the diffusion of propane in ZSM-5. This corroborates our previously reported work on the of influence of confinement on the rotational dynamics behavior of hydrocarbons (propane in particular) in ZSM-5 \cite{Dhiman}. Also, the diffusion coefficients obtained from the present study are comparable with the values obtained by Jobic et al. using QENS measurements, with values varying in the range of 12 $\times$ 10$^{-10}$ - 0.6 $\times$ 10$^{-10}$ m${^2}$s$^{-1}$ in NaZSM-5 with loading \cite{Jobic}. Reduction in diffusion coefficients with increased loading has been ascribed to molecular interactions, behavior similar to that observed in the present study. In comparison, the diffusion coefficient of propane in bulk is found at the most 1 orders of magnitude greater \cite{Schmid}, with values of 128 $\times$ 10$^{-10}$ m${^2}$s$^{-1}$ at 294 K, saturation pressure for bulk propane. 

The behavior of acetonitrile may be correlated with a high degree of charge asymmetry, which is in agreement with our previously reported study \cite{Dhiman}. A similar study by Hurle et al. on carbon disulphide and acetonitrile also indicates the important role of substantial dipole moment (charge asymmetry) in case of acetonitrile \cite{Hurle}.

Similarly, the molecular motion is observed to be greatly hindered for acetaldehyde. To the best of our knowledge not many studies focused on translation behavior, have been reported in literature. Nonetheless, it is equally interesting to compare propane and acetaldehyde, having moderately similar shape, and thus reiterates the important role played by charge distribution. This difference in charge distribution causes hindrance in the movement of acetaldehyde. 

In comparison to acetonitrile and acetaldehyde, acetone exhibits slightly less constrained movement in spite of the charge asymmetry. The globular shape of acetone might be responsible for this. Ertl et al. measured the self diffusion coefficient of bulk acetone, obtaining a value close to $\approx$ 40 $\times$ 10$^{-10}$ m${^2}$s$^{-1}$ \cite{Ertl}. This value is also in agreement with the one reported by Verros et al. \cite{Verros}. Also, Br${o}'$dka and Zerda performed molecular dynamics based studies on acetone under a comparatively less confined environment in silica, having cylindrically shaped pores of 1.5 - 3.0 nm diameters \cite{Brodka,Brodka1}. The diffusion coefficient decreases as a function of reducing pore size. Also in this study, the influence of electrostatic interactions on the diffusion coefficient is highlighted. As a result of these interactions, it is observed that motion of the molecules close to the surface, i. e., molecules residing within the contact layer of pores, become more restricted, while the movement of molecules in the center of the pore becomes faster. 

\section{\label{Conclusion} Conclusion}\

We have studied the effect of molecular shape on self-diffusion of guest molecules with similar kinetic diameters as a function of location within the three-dimensional structure of ZSM-5 zeolite framework. The ZSM-5 narrow channels produce a confinement effect and hinder the molecular motion - the effect of which differs from system to system. The different molecular shapes of the guest molecules gives rise to the charge asymmetry and consequently a net dipole moment except for propane. In this work, we have shown the effect of this phenomenon on the distribution of the molecules in different channels of ZSM-5 and on their dynamics both as a function of time and loading. It is evident that the non-linear shape and apolar nature of propane promotes occupancy in the intersections. A completely opposite behavior is observed for acetonitrile. Because of its linear shape and a net charge asymmetry, acetonitrile favors molecular motion in straight channels over others, with sinusoidal channel being the least preferred. The other two molecules which are non-linear with charge asymmetry mostly show similar behavior like acetronitrile. The consequence of this is clearly evident in their respective dynamics. The dynamics at the sub-picosecond level can be mainly ascribed to the ballistic regime where motion is collision free and independent of crowding effects. But beyond this domain, the confinement effect is distinctly observed based on molecular shape. Being the least hindered molecule, propane shows clear diffusive behavior with slower movement associated with increasing loading, while the motion of linear acetronitrile is restricted the most. The dynamical behavior of other two molecules, acetone and acetaldehyde, falls between propane and acetonitrile and tend to show a behavior close to diffusive at higher concentrations. Thus, in the present work, we have shown the influence of molecular shape on their motion in the confined environment ZSM-5. This work can also be used to estimate quantities that could be verified by experiments.         

\section{\label{Acknowledgement} Acknowledgement}\

This research was sponsored by the U.S. Department of Energy, Office of Science, Office of Basic Energy Sciences, under contract number DE-AC05-00OR22725 with UT-Battelle, LLC. DRC and SG would like to acknowledge support from the US Department of Energy, Office of Basic Energy Sciences, Division of Chemical Sciences, Geosciences and Biosciences, Geosciences Program under grant DE-SC0006878. 

non-commercial Creative Commons user license (CC-BY-NC-ND): ©\textless2018\textgreater. This manuscript version is made available under the CC-BY-NC-ND 4.0 license http://creativecommons.org/licenses/by-nc-nd/4.0/


\section*{\label{Ref} References}\

\end{document}